\documentclass[11pt]{article}

\usepackage{amsmath, amsthm, amssymb, fullpage, physics, algorithm, algorithmic, color, graphicx, url}

\title{Quantum Annealing Approaches to Solving the Shipment Rerouting Problems}

\author{Fei Li\thanks{Department of Computer Science, George Mason University, Fairfax, VA 22030. Email: fli4@gmu.edu} \and Arul Rhik Mazumder\thanks{School of Computer Science, Carnegie Mellon University, Pittsburgh, PA, 15213, USA. Email: arulm@andrew.cmu.edu} \and Max Zhao\thanks{Thomas Jefferson High School for Science and Technology, Alexandria, VA 22312, USA. Email: 2026mzhao@tjhsst.edu. This work was done when Max Zhao participated in the ASSIP program at George Mason University in the summer of 2024.}}

\date{}

\begin{document}

\maketitle

%%%%%%%%%%%%%%%%%%%%%%%%%%%%%%%%%%%%%%%%%%%%%%%%%%%%%%%%%%%%%

\begin{abstract}
In this paper, we study a shipment rerouting problem (SRP) which generalizes many NP-hard sequencing and packing problems. An SRP's solution has ample practical applications in vehicle scheduling and transportation logistics. Given a network of hubs, a set of goods must be delivered by trucks from their source-hubs to their respective destination-hubs. The objective is to select a set of trucks and to schedule these trucks' routes so that the total cost is minimized. The problem SRP is NP-hard; only classical approximation algorithms have been known for some of its NP-hard variants. In this work, we design classical algorithms and quantum annealing algorithms for this problem with various capacitated trucks. The algorithms that we design use novel mathematical programming formulations and new insights into solving sequencing and packing problems simultaneously. Such formulations take advantage of network infrastructure, shipments, and truck capacities. We conduct extensive experiments showing that in various scenarios, the quantum annealing solver generates near-optimal or optimal solutions much faster than the classical algorithm solver.
\end{abstract}

%%%%%%%%%%%%%%%%%%%%%%%%%%%%%%%%%%%%%%%%%%%%%%%%%%%%%%%%%%%%%

\section{Introduction}
\label{sec:intro}

We study a shipment rerouting problem that has been proposed in~\cite{OsabaRA24}. In this problem, there is a network that contains hubs and a set of transportation means (such as trucks or trains) that can have their runs on the network. For simplicity, we represent various transportation means using trucks with different capacities and different rental costs. A list of \emph{transportation requests}, which specify some amounts of \emph{goods} to be delivered from their sources to their respective destinations, should be satisfied. The objective is to minimize the total cost of successfully transporting these goods from their sources to their destinations.

%%%%%%%%%%%%%%%%%%%%%%%%%%%%%%%%%%%%%%%%%%%%%%%%%%%%%%%%%%%%%

\paragraph{Problem statement.}

The shipment rerouting problem is formulated as follows. We have a network represented by a directed graph $G = (V, E)$, where $V$ is the set of vertices and $E$ is the set of edges. Each edge $e \in E$ has a cost $c_e \ge 0$. A set of goods are to be transported by trucks with various capacities running over the network. A subset of vertices $H \subseteq V$ are labeled \emph{hubs} where goods can be loaded and unloaded. Note that a hub can be both some goods' source-hubs and some other goods' destination-hubs.

From the digraph $G$, we generate a complete digraph $F = (H, P)$ where the set of vertices $H \subseteq V$ denotes the hubs only and the set of edges $p(i, j) \in P$ denote the shortest paths connecting two hubs $i \in H$ and $j \in H$ in the transportation network $G = (V, E)$ -- these path costs $c({p(i, j)}) := \sum_{e \in p(i, j)} c_e$ can be precalculated using $|V|$ rounds of the Dijkstra's algorithm~\cite{CormenLRS22} in time $O(|E||V| + |V|^2 \log |V|)$. The generated digraph $F$ has its edge costs satisfy the triangular inequalities. Consider the complete digraph $F = (H, P)$. Let $|H| = n$ and we label these hubs $\in H$ as $1, 2, \ldots, n$. There are $m$ \emph{transportation requests} of truck runs for transporting goods. We use a triple $(s_i, t_i, \ell_i)$ to denote a truck run $i$ ($i = 1, 2, \ldots, m$), where $s_i$ is the source-hub, $t_i$ is the destination-hub, and $\ell_i$ is the goods' load. There are two constraints:
\begin{itemize}
    \item ($A_1$) The goods cannot be divided into fractions during transportation.
    
    \item ($A_2$) Once some goods are loaded on a truck, these goods cannot be unloaded before the truck reaches these goods' destination. That is, goods can be loaded only at these goods' source-hubs and unloaded at their destination-hubs.
\end{itemize}

There are a set of trucks with various capacities. For goods with load $\ell_i$, a truck with its remaining capacity $\ge \ell_i$ can load the goods at $s_i$ and unload the goods at $t_i$. There is a list of $K$ trucks (transportation means) with their capacities $L_1, L_2, \ldots, L_K$ respectively, and for these $K$ trucks, we pay costs $R_1, R_2, \cdots, R_K$ respectively for renting and operating them. Without loss of generality, we have two assumptions: ($1$) $L_1 \le L_2 \le \ldots \le L_K$ and ($2$) $R_1 \le R_2 \le \cdots \le R_K$. All the values (edge costs, loads of goods, truck capacities, rental fees) are positive numbers. Subject to both constraints $A_1$ and $A_2$, the objective is to minimize the total cost of truck rental fees and $C$ times the traveling distances of trucks transporting all the goods from their sources to their respective destinations, where $C > 0$ is a positive constant representing the money spent per unit of traveling distance. We name this problem SRP (shipment rerouting problem).

%%%%%%%%%%%%%%%%%%%%%%%%%%%%%%%%%%%%%%%%%%%%%%%%%%%%%%%%%%%%%

\paragraph{Related work.}

Many variants of the shipment rerouting problem SRP have been studied in the past. The problem SRP is intractable, as shown below. Consider a setting in which all the trucks have capacities more than the total goods' load $\sum_i \ell_i$ and the minimum rental cost $R_1$ is more than the total cost of all the paths among all the hubs. In addition, assume that all the transportation requests are very far apart from each other compared to the traveling distance of each transportation request. Thus, to minimize the total cost paid to transport all the goods, we only need one truck to ship all the goods, and the optimal solution to this SRP's instance minimizes the total traveling distance by visiting all the hubs. The special instance of SRP is reduced to a \emph{NP-complete sequencing problem}, the traveling salesman problem~\cite{LawlerLKS85}. Now, we consider another special case in which all the transportation requests are with the same source-hub and the same destination-hub. In addition, the network is just a line graph. We show that this variant of SRP is NP-complete. Consider $n$ transportation requests, each transportation request $i$ having a need $\ell_i$. There are only two identical trucks, each having a capacity equal to $\sum \ell_i / 2$. Thus, we need to decide which request is assigned to which truck to carry from the source to the destination. This problem reduces to an \emph{NP-complete packing problem}, the partition problem~\cite{GareyJ79}.

To our knowledge, there are no classical approximation algorithms known for this shipment rerouting problem. Part of the reason is that an algorithm's approximation ratio is very sensitive to the values in the input instance. It is easy to construct an instance so that an approximation ratio can be changed from $1$ to an arbitrary number given a slight change in the shipment loads, the truck capacities, and/or the rental fees. For such a problem, we consider not only how to schedule shipments to minimize the total traveling distance but also how to pack goods to satisfy the truck capacity constraints. A single algorithmic technique generally cannot optimize the objective under both constraints.

The most recent work related to ours is~\cite{OsabaRA24}. In this work~\cite{OsabaRA24}, Osaba \emph{et al.} gave a quadratic unconstrained binary optimization (QUBO) formulation to solve the problem. However, this QUBO-based formulation works only for a single truck route, and their algorithm is to iteratively select truck routes one by one to serve all transportation requests. This heuristic approach is not guaranteed to generate an optimal set of trucks to deliver goods, and no proof shows its optimality~\cite{OsabaRA24}. The QUBO formulation is solved using an Ising model-based quantum annealing approach~\cite{Lucas14} in quantum computers.

Note that if we remove the above constraint $A_2$, then the goods are allowed to be unloaded from one truck at some hubs and to be loaded by another truck at the same hubs before they reach their destination. We name this problem a \emph{segmented shipment rerouting problem (SSRP)}. The model studied in~\cite{YarkoniHSSTLBN21} is a special case of SSRP in which all the trucks have identical capacities and the same rental cost of $0$ and the objective is to minimize the total traveling distances. In~\cite{YarkoniHSSTLBN21}, Yarkoni \emph{et. al} formulated the special case of SSRP as a QUBO problem and used quantum computers and classic computers to calculate the solutions. However, in their formulation, the number of variables (on the number of candidate paths for transporting a good) can be exponentially large.

Another related work in designing logistic networks is the problem studied in~\cite{DingCLSS21}, which is the same as the facility location problem~\cite{Vazirani03, WilliamsonS11}. Ding \emph{et al.}~\cite{DingCLSS21} solved the problem using a quantum annealing approach. The constraints are different from the ones in~\cite{YarkoniHSSTLBN21}, and so are the formulations. Neither~\cite{YarkoniHSSTLBN21} nor~\cite{DingCLSS21} considered trucks with different capacities and rental fees. Note that allowing trucks to have various capacities makes scheduling and packing more challenging.

%%%%%%%%%%%%%%%%%%%%%%%%%%%%%%%%%%%%%%%%%%%%%%%%%%%%%%%%%%%%%

\paragraph{Our contributions.}

In this paper, we design classical algorithms and quantum algorithms for the shipment rerouting problem SRP. From the algorithm design perspective, our contributions include ($1$) two mixed integer program formulations that generate exact solutions to SRP, and ($2$) two quantum annealing algorithms based on the constrained quadratic model (CQM) that generate solutions to SRP. From an empirical study perspective, we study the performance of our algorithms on various network topologies and transportation requests. The experimental results guide us in selecting the quantum annealing algorithms for the shipment rerouting problem in terms of solution qualities and running time.

%%%%%%%%%%%%%%%%%%%%%%%%%%%%%%%%%%%%%%%%%%%%%%%%%%%%%%%%%%%%%

\paragraph{Paper organization.}

In Section~\ref{sec:classical_srp}, we introduce our classical algorithmic solutions to the shipment rerouting problem SRP. In Section~\ref{sec:added}, we present our quantum annealing solutions. In Section~\ref{sec:experiments}, we conduct extensive experiments to compare our classical and quantum solutions in various settings. We conclude our paper in Section~\ref{secc:conclusion}.

%%%%%%%%%%%%%%%%%%%%%%%%%%%%%%%%%%%%%%%%%%%%%%%%%%%%%%%%%%%%%

\section{Classical Algorithms}
\label{sec:classical_srp}

%%%%%%%%%%%%%%%%%%%%%%%%%%%%%%%%%%%%%%%%%%%%%%%%%%%%%%%%%%%%%

In this section, we introduce two mixed linear programs that generate exact solutions to the shipment rerouting problem SRP, without and with a dispatch center.

Consider a complete digraph $F = (H, P)$ with $|H| = n$. Consider $m$ transportation requests $(s_1, t_1, \ell_1), (s_2, t_2, \ell_2), \ldots, (s_m, t_m, \ell_m)$. Consider $K$ trucks and we label them as $1, 2, \ldots, K$. These $K$ trucks have capacities $L_1 \le L_2 \le \cdots \le L_K$ and their corresponding rental fees are $R_1, R_2, \ldots, R_K$ respectively. We use $d(p(a, b))$ to denote the distance between a hub $a$ and a hub $b$. We have two constraints in SRP:
\begin{itemize}
    \item ($A_1$) The goods cannot be divided into fractions during transportation.
    
    \item ($A_2$) The goods can be loaded only at their respective source-hubs and unloaded only at their respective destination-hubs.
\end{itemize}

Our following solution generates some truck routes to satisfy all transportation requests at the minimum cost. A truck route must start from one source-hub and end at one destination-hub. We only list the hubs that a truck must visit in the graph $F = (H, P)$. A truck visits a hub if and only if it needs to load or unload some goods at that hub. A truck's route has at most $2 m$ possible \emph{stops} since a truck can visit at most $m$ source-hubs and $m$ destination-hubs. For any truck, we use $1, 2, \ldots, 2 m$ to index the \emph{stops} that the truck may visit. We note here that, from the notation of stops, if a truck makes multiple times of loading and unloading goods at the same hub, then we have multiple distinctly indexed stops representing these loading/unloading spots on this truck's route though these stops share the same hub.

We define the indicator variables in Table~\ref{tab:var} and present the conditions that need to be satisfied over these $0$-$1$ variables in the list from $C_1$ to $C_8$. The objective is listed as $C_9$.

\begin{table}[h!]
    \centering
    \begin{tabular}{l|p{5.6in}}
    \hline
    $X(i, j, p)$ &  an indicator variable showing whether or not the request $(s_i, t_i, \ell_i)$'s source $s_i$ appears as the $p$-th stop on the truck $j$'s route,
    \begin{displaymath}
    X(i, j, p) = \begin{cases}
    1, & \mbox{if $s_i$ is at the $p$th spot on the truck $j$'s route}\\
    0, & \mbox{otherwise}
    \end{cases}
    \end{displaymath}\\ \hline
    $Y(i, j, p)$ &  an indicator variable showing whether or not the request $(s_i, t_i, \ell_i)$'s destination $t_i$ appears as the $p$-th stop on the truck $j$'s route,
    \begin{displaymath}
    Y(i, j, p) = \begin{cases}
    1, & \mbox{if $t_i$ is at the $p$th spot on the truck $j$'s route}\\
    0, & \mbox{otherwise}
    \end{cases}
    \end{displaymath}\\ \hline
    $Z(j)$ & an indicator variable showing whether or not the truck $j$ is selected to serve one or more transportation requests.
    \begin{displaymath}
    Z(j) = \begin{cases}
    1, & \mbox{if the truck $j$ is selected to serve transportation requests}\\
    0, & \mbox{otherwise}
    \end{cases}
    \end{displaymath}
    \\ \hline
    \end{tabular}
    \caption{$0$-$1$ variables used in the integer linear program for SRP}
    \label{tab:var}
\end{table}

\begin{itemize}
    \item[$C_1$.] A transportation request must be served by some truck.
    
    The source and the destination of a transportation request must be placed at some stops of a truck's route.

    \begin{align*}
    \sum^K_{j = 1} \sum^{2m}_{p = 1} X(i, j, p) = 1, & & i = 1, \ldots, m\\
    \sum^K_{j = 1} \sum^{2m}_{p = 1} Y(i, j, p) = 1, & & i = 1, \ldots, m
    \end{align*}

    \item[$C_2$.] For each stop $p$ on a route, we have at most one transportation request being served (either loading or unloading goods).

    \begin{align*}
    \sum^{m}_{i = 1} \left[X(i, j, p) + Y(i, j, p)\right] & \le 1, & j = 1, \ldots, K, \ p = 1, \ldots, 2m
    \end{align*}
    
    \item[$C_3$.] A transportation request has its source-hub and destination-hub on the same route (due to the constraint $A_2$).

    \begin{align*}
    \sum^{2m}_{p = 1} X(i, j, p) = \sum^{2m}_{p = 1} Y(i, j, p), & & i = 1, \ldots, m, \ j = 1, \ldots, K 
    \end{align*}

    \item[$C_4$.] A request source-hub should be placed at a stop before its destination-hub on the same route.
    
    This condition, when $q = 2 m$ holds, is replaced by the above condition $C_3$ in our formulation.

    \begin{align*}
    \sum^q_{p = 1} Y(i, j, p) \le \sum^q_{p = 1} X(i, j, p), & & i = 1, \ldots, m, \ j = 1, \ldots, K, \ q = 1, \ldots, 2 m
    \end{align*}

    \item[$C_5$.] A truck is chosen if it has transportation requests to be served on its route.

    \begin{align*}
    Z(j) & \ge \sum^{2m}_{p = 1} X(i, j, p), & & i = 1, \ldots, m, \ j = 1, \ldots, K\\
    Z(j) & \ge \sum^{2m}_{p = 1} Y(i, j, p), & & i = 1, \ldots, m, \ j = 1, \ldots, K
    \end{align*}

    \item[$C_6$.] A truck on a route should have its capacity no less than the load it carries.
    
    Consider a truck $j$. At the $q$-th stop, if $\sum^q_{p = 1} X(i, j, p) - \sum^q_{p = 1} Y(i, j, p) = 1$, then the truck $j$ is carrying the load $\ell_i$ of the transportation request $i$. Otherwise (i.e., if $\sum^q_{p = 1} X(i, j, p) - \sum^q_{p = 1} Y(i, j, p) = 0$), then the truck is not carrying the load $\ell_i$ at the $q$-th stop.

    \begin{align*}
    L_j & \ge \sum^m_{i = 1} \left[\left(\sum^q_{p = 1} X(i, j, p) - \sum^q_{p = 1} Y(i, j, p)\right) \cdot \ell_i\right], & j = 1, \ldots, K, \ q = 1, \ldots, 2m
    \end{align*}

    We remark here that if there exist multiple types of capacity constraints such as the truck's size constraint or the truck's weight constraint, we then apply the same inequality (replacing $L_j$ and $\ell_i$ using the other type constraints' variables) to generate the conditions that need to be satisfied.
    
    \item[$C_7$.] A truck, if used to transport goods, should serve a transportation request by loading it at its first stop.

    \begin{align*}
    \sum^m_{i = 1} X(i, j, 1) = Z(j), & & j = 1, 2, \ldots, K
    \end{align*}

    Satisfying this condition $C_7$ and the following condition $C_8$ helps us correctly calculate the total traveling distance for the truck $j$ in the objective $C_9$. 
    
    \item[$C_8$.] A truck should serve the transportation requests consecutively along its stops.

    The indicator variables $X(i, j, p)$ and $Y(i, j, p)$, if they are non-zero, should indicate that a truck loads and unloads goods consecutively along its stops. That is, if a stop $p$ in a truck $j$'s route is not used to load or unload goods (i.e., if $\sum^m_{i = 1} X(i, j, p) = 0$ and $\sum^m_{i = 1} Y(i, j, p) = 0$), then the following stops (the $(p + 1)$th stop, the $(p + 2)$th stop, $\ldots$, the $(2m)$th stop) in the same route should not be used to load or unload goods (i.e., $\sum^m_{i = 1} X(i, j, p') = 0$ and $\sum^m_{i = 1} Y(i, j, p') = 0$, where $p' = p + 1, p + 2, \ldots, 2m$). To guarantee this condition $C_8$ to hold, we add the following inequalities.

    \begin{align*}
    & \sum^m_{i = 1} \left[X(i, j, p - 1) + Y(i, j, p - 1)\right] + \sum^m_{i = 1} \left[X(i, j, p) + Y(i, j, p)\right] \ge 2 \sum^m_{i = 1} \left[X(i, j, p + 1) + Y(i, j, p + 1)\right]\\
    & j = 1, \ldots, K, \ p = 2, \ldots, 2m - 1
    \end{align*}
    
    \item[$C_9$.] The objective is to minimize the total rental fee and $C$ times of the total traveling cost.

    For any two neighboring stops (the $(q - 1)$th stop and the $q$th stop) on the truck $j$'s route, we count the cost $w_{j, q}$ to the truck $j$'s traveling distance, where $w_{j, q}$ is calculated below.
    
    \begin{displaymath}
    w_{j, q} = \sum^m_{i = 1, i \neq i'} \sum^m_{i' = 1}
    \begin{cases}
    d(p(s_i, s_{i'})) & \mbox{if } X(i, j, q - 1) = X(i', j, q)  = 1\\
    d(p(t_i, s_{i'})) & \mbox{if } Y(i, j, q - 1) = X(i', j, q)  = 1\\
    d(p(s_i, t_{i'})) & \mbox{if } X(i, j, q - 1) = Y(i', j, q) = 1\\
    d(p(t_i, t_{i'})) & \mbox{if } Y(i, j, q - 1) = Y(i', j, q) = 1
    \end{cases}
    \end{displaymath}

    Thus, $w_{j, q}$ is rewritten as

    \begin{multline*}
    w_{j, q} = \sum^m_{i = 1} \sum^m_{i' = 1} \left(d(p(s_i, s_{i'})) \left\lceil \frac{ X(i, j, q - 1) + X(i', j, q)  - 1}{2} \right\rceil \right.\\
    \left. + d(p(t_i, s_{i'})) \left\lceil \frac{Y(i, j, q - 1) + X(i', j, q)  - 1}{2} \right\rceil +  d(p(s_i, t_{i'})) \left\lceil \frac{ X(i, j, q - 1) + Y(i', j, q) - 1}{2} \right\rceil \right.\\
    \left. + d(p(t_i, t_{i'})) \left\lceil \frac{Y(i, j, q - 1) + Y(i', j, q) - 1}{2} \right\rceil \right)\\
    j = 1, 2, \ldots, K, \ q = 2, 3, \ldots, 2m
    \end{multline*}

    The objective is the following:
    \begin{eqnarray*}\\
    \min & \sum^K_{j = 1} R_j \cdot Z(j) + C \sum^K_{j = 1} \sum^{2 m}_{q = 2} w_{j, q}
    \end{eqnarray*}
    where $C$ is the constant monetary cost spent per unit of traveling distance.
\end{itemize}

In the above formulations, we optimize the sum of the total rental fee and $C$ times the total cost of transporting all the goods from their source-hubs to their respective destination-hubs. The total number of variables ($X(i, j, p), Y(i, j, p), Z(j), w_{j, q}$) is $m \cdot K \cdot 2m \cdot 2 + K + K \cdot 2m = O(m^2 K)$. The total number of conditions is $4 m^2 K + 2m + K \cdot 2m + m \cdot K + m \cdot K \cdot 2m + 2 m \cdot K + 2 m \cdot K + 2m \cdot K + 2m \cdot K = O(m^2 K)$.

%%%%%%%%%%%%%%%%%%%%%%%%%%%%%%%%%%%%%%%%%%%%%%%%%%%%%%%%%%%%%

\paragraph{A mixed integer program formulation with a dispatch center.}

The above formulation ($C_1$ to $C_9$) does not enforce that all trucks start from one dispatch center and return to the same dispatch center after their runs. In the following, we assume that there is a dispatch center, denoted by $O$. We need to dispatch all trucks from $O$ to serve all transportation requests, and these trucks are required to get back to $O$ after their service. We thus modify the objective by adding the total traveling cost $c_\alpha$ from the dispatch center $O$ to the first stop of those trucks serving requests and the total traveling cost $c_\beta$ from the last stop of serving requests back to the dispatch center.

For $c_\alpha$, we have
\begin{displaymath}
c_\alpha = \sum^K_{j = 1} \sum^m_{i = 1} \left[d(p(O, s_i)) \cdot X(i, j, 1)\right]
\end{displaymath}
where $d(p(O, s_i))$ denotes the shortest path length from the dispatch center $O$ to the source-hub $s_i$ of the request $(s_i, t_i, \ell_i)$.

For $c_\beta$, we take the following approach. From the constraint $C_8$, a truck serves its transportation requests consecutively. We thus identify the destination-hub of the last request served by examining a stop and its immediately following stop. To make sure that any truck's route has a stop without loading or unloading goods, we add a placeholder at the end of each truck's last stop and name it the $(2m + 1)$th stop. We also enforce $X(i, j, 2m + 1) = Y(i, j, 2m + 1) = 0$, where $i = 1, \ldots, m$ and $j = 1, \ldots, K$ to ensure that no loading or unloading occurs at the $(2m + 1)$th stop. This placeholder helps us identify the transportation request being served as the last one for a given truck. If the destination-hub $t_i$ of the request $(s_i, t_i, \ell_i)$ is at the last stop for a truck $j$ to unload goods $i$, let this stop be $j$'s $q$-th stop and we thus have
\begin{align*}
Y(i, j, q) & = 1\\
\sum^m_{i = 1} X(i, j, q + 1) & = \sum^m_{i = 1} Y(i, j, q + 1) = 0
\end{align*}

For any other possible hub $q'$ ($q' \neq q$) on the truck $j$'s route, we have
\begin{align*}
Y(i, j, q') & = 0\\
\sum^m_{i = 1} X(i, j, q' + 1) + \sum^m_{i = 1} Y(i, j, q' + 1) & = 1 \mbox{ or } 0
\end{align*}

Thus, we identify that the truck $j$ serves the transportation request $i$ in $j$'s last stop by having
\begin{align*}
\sum^{2m}_{q = 1} \left\lceil\frac{Y(i, j, q) - \sum^m_{i' = 1} \left[X(i', j, q + 1) + Y(i', j, q + 1)\right]}{2}\right\rceil & = \\
&
\begin{cases}
1, & \mbox{ $i$ is the last request served by the truck $j$}\\
0, & \mbox{ $i$ is not the last request served by the truck $j$}
\end{cases}
\end{align*}

After identifying the last transportation request $i$ served by the truck $j$, we calculate the distance from its destination-hub back to the dispatch center.

\begin{displaymath}
c_\beta = \sum^K_{j = 1} \sum^m_{i = 1} \sum^{2m}_{q = 1} \left\lceil\frac{Y(i, j, q) - \sum^m_{i' = 1} \left(X(i', j, q + 1) + Y(i', j, q + 1)\right)}{2}\right\rceil d(p(t_i, O))
\end{displaymath}
where $d(p(t_i, O))$ denotes the shortest path length from the destination-hub $t_i$ of the request $(s_i, t_i, \ell_i)$ to the dispatch center $O$.

The objective of the SRP with a dispatch center then becomes
\begin{eqnarray*}
\min & c_\alpha + c_\beta + \sum^K_{j = 1} R_j \cdot Z(j) + C \sum^K_{j = 1} \sum^{2 m}_{q = 2} w_{j, q}
\end{eqnarray*}

The total number of variables ($X(i, j, p), Y(i, j, p), Z(j), w_{j, q}$) is $m \cdot K \cdot 2m \cdot 2 + K + K \cdot 2m = O(m^2 K)$. The total number of conditions is $4 m^2 K + 2m + K \cdot 2m + m \cdot K + m \cdot K \cdot 2m + 2 m \cdot K + 2 m \cdot K + 2m \cdot K + 2m \cdot K = O(m^2 K)$.

%%%%%%%%%%%%%%%%%%%%%%%%%%%%%%%%%%%%%%%%%%%%%%%%%%%%%%%%%%%%%

%%%%%%%%%%%%%%%%%%%%%%%%%%%%%%%%%%%%%%%%%%%%%%%%%%%%%%%%%%%%%

\section{Quantum Annealing Algorithms}
\label{sec:added}

%%%%%%%%%%%%%%%%%%%%%%%%%%%%%%%%%%%%%%%%%%%%%%%%%%%%%%%%%%%%%

In this section, we introduce two formulations based on the constrained quadratic model (CQM) to generate solutions to the shipment rerouting problem SRP, without and with a dispatch center. The CQM solver can model problems defined on binary and integer values. The CQM solver provides direct support for expressing a variety of constraints, including variable bounds. linear and quadratic equalities.

%Consider a complete digraph $F = (H, P)$ with $|H| = n$. Consider $m$ transportation requests $(s_1, t_1, \ell_1), (s_2, t_2, \ell_2), \ldots, (s_m, t_m, \ell_m)$. Consider $K$ trucks and we label them as $1, 2, \ldots, K$. These $K$ trucks have capacities $L_1 \le L_2 \le \cdots \le L_K$ and their corresponding rental fees are $R_1, R_2, \ldots, R_K$ respectively. We use $c(p(a, b))$ to denote the distance between a hub $a$ and a hub $b$. We have two constraints in SRP:
%\begin{itemize}
%    \item ($A_1$) The goods cannot be divided into fractions during transportation.
    
%    \item ($A_2$) The goods can be loaded only at their respective source-hubs and unloaded only at their respective destination-hubs.
%\end{itemize}

Our following solution generates some truck routes to satisfy all transportation requests at the minimum cost. A truck route must start from one source-hub and end at one destination-hub. We only list the hubs that a truck must visit in the graph $F = (H, P)$. A truck visits a hub if and only if it needs to load or unload some goods at that hub. A truck's route has at most $2 m$ possible \emph{stops} since a truck can visit at most $m$ source-hubs and $m$ destination-hubs. For any truck, we use $1, 2, \ldots, 2 m$ to index the \emph{stops} that the truck may visit. We define the same set of indicator variables as in Table~\ref{tab:var} and present the conditions that need to be satisfied over these $0$-$1$ variables in the list from $D_1$ to $D_8$. The objective is listed as $D_9$.

\begin{itemize}
    \item[$D_1$.] A transportation request must be served by some truck.

    \begin{align*}
    \sum^K_{j = 1} \sum^{2m}_{p = 1} X(i, j, p) = 1, & & i = 1, \ldots, m\\
    \sum^K_{j = 1} \sum^{2m}_{p = 1} Y(i, j, p) = 1, & & i = 1, \ldots, m
    \end{align*}

    \item[$D_2$.] For each stop $p$ on a route, we have at most one transportation request being served (either loading or unloading goods).

    \begin{align*}
    \sum^{m}_{i = 1} \left[X(i, j, p) + Y(i, j, p)\right] & \le 1, & j = 1, \ldots, K, \ p = 1, \ldots, 2m
    \end{align*}
    
    \item[$D_3$.] A transportation request has its source-hub and destination-hub on the same route (due to the constraint $A_2$).

    \begin{align*}
    \sum^{2m}_{p = 1} X(i, j, p) = \sum^{2m}_{p = 1} Y(i, j, p), & & i = 1, \ldots, m, \ j = 1, \ldots, K 
    \end{align*}

    \item[$D_4$.] A request source-hub should be placed at a stop before its destination-hub on the same route.
    
    This condition, when $q = 2 m$ holds, is replaced by the above condition $D_3$ in our formulation.

    \begin{align*}
    \sum^q_{p = 1} Y(i, j, p) \le \sum^q_{p = 1} X(i, j, p), & & i = 1, \ldots, m, \ j = 1, \ldots, K, \ q = 1, \ldots, 2 m
    \end{align*}

    \item[$D_5$.] A truck is chosen if it has transportation requests to be served on its route.

    \begin{align*}
    Z(j) & \ge \sum^{2m}_{p = 1} X(i, j, p), & & i = 1, \ldots, m, \ j = 1, \ldots, K\\
    Z(j) & \ge \sum^{2m}_{p = 1} Y(i, j, p), & & i = 1, \ldots, m, \ j = 1, \ldots, K
    \end{align*}

    \item[$D_6$.] A truck on a route should have its capacity no less than the load it carries.

    \begin{align*}
    L_j & \ge \sum^m_{i = 1} \left[\left(\sum^q_{p = 1} X(i, j, p) - \sum^q_{p = 1} Y(i, j, p)\right) \cdot \ell_i\right], & j = 1, \ldots, K, \ q = 1, \ldots, 2m
    \end{align*}

    \item[$D_7$.] A truck, if used to transport goods, should serve a transportation request by loading it at its first stop.

    \begin{align*}
    \sum^m_{i = 1} X(i, j, 1) = Z(j), & & j = 1, 2, \ldots, K
    \end{align*}
    
    \item[$D_8$.] A truck should serve the transportation requests consecutively along its stops.

    The indicator variables $X(i, j, p)$ and $Y(i, j, p)$, if they are non-zero, should indicate that a truck loads and unloads goods consecutively along its stops. That is, if a stop $p$ in a truck $j$'s route is not used to load or unload goods (i.e., if $\sum^m_{i = 1} X(i, j, p) = 0$ and $\sum^m_{i = 1} Y(i, j, p) = 0$), then the following stops (the $(p + 1)$th stop, the $(p + 2)$th stop, $\ldots$, the $(2m)$th stop) in the same route should not be used to load or unload goods (i.e., $\sum^m_{i = 1} X(i, j, p') = 0$ and $\sum^m_{i = 1} Y(i, j, p') = 0$, where $p' = p + 1, p + 2, \ldots, 2m$). To guarantee this condition $D_8$ to hold, we add the following inequalities.

    \begin{eqnarray*}
    & & \left(\sum^m_{i = 1} \left[X(i, j, p - 1) + Y(i, j, p - 1)\right]\right) \left(\sum^m_{i = 1} \left[X(i, j, p) + Y(i, j, p)\right]\right)\\
    & \ge & \sum^m_{i = 1} \left[X(i, j, p + 1) + Y(i, j, p + 1)\right]\\
    & & j = 1, \ldots, K, \ p = 2, \ldots, 2m - 1
    \end{eqnarray*}
    
    \item[$D_9$.] The objective is to minimize the total rental fee and the total traveling cost.

    For any two neighboring stops $q - 1$ and $q$ on the truck $j$'s route, we count the cost $w_{j, q}$ that the truck $j$'s traveling distance, where $w_{j, q}$ is calculated below,
    \begin{displaymath}
    w_{j, q} = \sum^m_{i = 1, i \neq i'} \sum^m_{i' = 1}
    \begin{cases}
    c(p(s_i, s_{i'})) & \mbox{if } X(i, j, q - 1) \cdot X(i', j, q)  = 1\\
    c(p(t_i, s_{i'})) & \mbox{if } Y(i, j, q - 1) \cdot X(i', j, q)  = 1\\
    c(p(s_i, t_{i'})) & \mbox{if } X(i, j, q - 1) \cdot Y(i', j, q) = 1\\
    c(p(t_i, t_{i'})) & \mbox{if } Y(i, j, q - 1) \cdot Y(i', j, q) = 1
    \end{cases}
    \end{displaymath}

    Thus, $w_{j, q}$ is rewritten as

    \begin{multline*}
    w_{j, q} = \sum^m_{i = 1} \sum^m_{i' = 1} c(p(s_i, s_{i'})) \cdot X(i, j, q - 1) \cdot X(i', j, q) + c(p(t_i, s_{i'})) \cdot Y(i, j, q - 1) \cdot X(i', j, q)\\
    \left. +  c(p(s_i, t_{i'})) \cdot X(i, j, q - 1) \cdot Y(i', j, q) + c(p(t_i, t_{i'})) \cdot Y(i, j, q - 1) \cdot Y(i', j, q) \right.\\
    \left. j = 1, 2, \ldots, K \ q = 2, 3, \ldots, 2m \right.
    \end{multline*}

    The objective is the following:
    \begin{eqnarray*}\\
    \min & \sum^K_{j = 1} R_j \cdot Z(j) + C \sum^K_{j = 1} \sum^{2 m}_{q = 2} w_{j, q}
    \end{eqnarray*}
    where $C$ is the constant monetary cost spent per unit of traveling distance.
\end{itemize}

We now formulate the quantum annealing solution for the problem SRP with a dispatch center $O$. Let $c_\alpha$ denote the total traveling cost from the dispatch center $O$ to the first stop of those trucks serving requests and $c_\beta$ the total traveling cost from the last stop of serving requests back to the dispatch center.

For $c_\alpha$, we have
\begin{displaymath}
c_\alpha = \sum^K_{j = 1} \sum^m_{i = 1} \left[d(p(O, s_i)) \cdot X(i, j, 1)\right]
\end{displaymath}
where $d(p(O, s_i))$ denotes the shortest path length from the dispatch center $O$ to the source-hub $s_i$ of the request $(s_i, t_i, \ell_i)$.

For $c_\beta$, we take the following approach. To make sure that any truck's route has a stop without loading or unloading goods, we add a placeholder at the end of each truck's last stop and name it the $(2m + 1)$th stop. We also enforce $X(i, j, 2m + 1) = Y(i, j, 2m + 1) = 0$, where $i = 1, \ldots, m$ and $j = 1, \ldots, K$ to ensure that no loading or unloading occurs at the $(2m + 1)$th stop. This placeholder helps us identify the transportation request being served as the last one for a given truck. If the destination-hub $t_i$ of the request $(s_i, t_i, \ell_i)$ is at the last stop for a truck $j$ to unload goods $i$, let this stop be $j$'s $q$-th stop and we thus have
\begin{align*}
Y(i, j, q) & = 1\\
\sum^m_{i = 1} X(i, j, q + 1) & = \sum^m_{i = 1} Y(i, j, q + 1) = 0
\end{align*}

For any other possible hub $q'$ ($q' \neq q$) on the truck $j$'s route, we have
\begin{align*}
Y(i, j, q') & = 0\\
\sum^m_{i = 1} X(i, j, q' + 1) + \sum^m_{i = 1} Y(i, j, q' + 1) & = 1 \mbox{ or } 0
\end{align*}

Thus, we identify that the truck $j$ serves the transportation request $i$ in $j$'s last stop by having
\begin{align*}
\sum^{2m}_{q = 1} Y(i, j, q) \left(1 - \sum^m_{i' = 1} \left[X(i', j, q + 1) + Y(i', j, q + 1)\right]\right) & = \\
&
\begin{cases}
1, & \mbox{ $i$ is the last request served by the truck $j$}\\
0, & \mbox{ $i$ is not the last request served by the truck $j$}
\end{cases}
\end{align*}

After identifying the last transportation request $i$ served by the truck $j$, we calculate the distance from its destination-hub back to the dispatch center.

\begin{displaymath}
c_\beta = \sum^K_{j = 1} \sum^m_{i = 1} \sum^{2m}_{q = 1} Y(i, j, q) \left(1 - \sum^m_{i' = 1} \left[X(i', j, q + 1) + Y(i', j, q + 1)\right]\right) d(p(t_i, O))
\end{displaymath}
where $d(p(t_i, O))$ denotes the shortest path length from the destination-hub $t_i$ of the request $(s_i, t_i, \ell_i)$ to the dispatch center $O$.

The objective of the SRP with a dispatch center then becomes
\begin{eqnarray*}
\min & c_\alpha + c_\beta + \sum^K_{j = 1} R_j \cdot Z(j) + C \sum^K_{j = 1} \sum^{2 m}_{q = 2} w_{j, q}
\end{eqnarray*}

The total number of variables ($X(i, j, p), Y(i, j, p), Z(j), w_{j, q}$) is $m \cdot K \cdot 2m \cdot 2 + K + K \cdot 2m = O(m^2 K)$. The total number of conditions is $4 m^2 K + 2m + K \cdot 2m + m \cdot K + m \cdot K \cdot 2m + 2 m \cdot K + 2 m \cdot K + 2m \cdot K + 2m \cdot K = O(m^2 K)$.

%%%%%%%%%%%%%%%%%%%%%%%%%%%%%%%%%%%%%%%%%%%%%%%%%%%%%%%%%%%%%

\section{Experiments}
\label{sec:experiments}

In this section, we design experiments to compare the classical algorithms and the quantum algorithms on real-life data sets. We name our classical algorithms SRP (without a dispatch center) and SRP-D (with a dispatch center). We name our quantum annealing algorithms QA (without a dispatch center) and QA-D (with a dispatch center).

%%%%%%%%%%%%%%%%%%%%%%%%%%%%%%%%%%%%%%%%%%%%%%%%%%%%%%%%%%%%%

\subsection{Settings}

\paragraph{On input data.}

The input data instance comes from Transportation Networks Github~\cite{Stabler_Bar-Gera_Sal}. This data repository was initially developed to solve the traffic assignment problem~\cite{Patriksson2015TheTA}, which is very similar to the shipment routing problem. We tested the algorithms in Section~\ref{sec:classical_srp} and Section~\ref{sec:added} across $m = 1, 2, 3, 4, 5$ shipments, each on the following $6$ networks: \emph{Sioux Falls}, \emph{Eastern Massachusetts}, \emph{Anaheim}, \emph{Chicago}, \emph{Barcelona}, and \emph{Winnipeg} described in Table~\ref{tab:graphs}. 

\begin{table}[h!]
    \centering
    \begin{tabular}{|l|r|r|}
        \hline
        names & \# of nodes & \# of edges \\ \hline \hline
        Sioux Falls & 24 & 76 \\ \hline
        Eastern Massachusetts & 74 & 258 \\ \hline
        Anaheim & 416 & 914  \\ \hline
        Chicago & 933 & 2950 \\ \hline
        Barcelona & 1020 & 2552 \\ \hline
        Winnipeg & 1052 & 2836 \\ \hline
    \end{tabular}
    \caption{Testbed: $5$ networks}
    \label{tab:graphs}
\end{table}

Given the underlying network, the $m$ shipments are randomly generated so that they do not share the same source $s_i$ and destination $t_i$ hub. Each shipment is assumed to carry a fixed constant load of $\ell_i = 50$. We also assume that we have $K = m$ trucks available. These trucks are also assumed to have a constant fixed capacity of $L_i =100$ and a fixed rental fee of $R_{i} = 1$. We also set a fixed cost of $C = 1$ per unit distance traveled.

We perform the tests using $m = 1, 2, 3, 4, 5$ due to the variable and constraint scaling. After $m = 5$, we reach a bottleneck in terms of the number of variables and constraints allowable for both the CPLEX Branch-and-Cut Solver and the D-Wave's LeapHybridCQMSampler.

%%%%%%%%%%%%%%%%%%%%%%%%%%%%%%%%%%%%%%%%%%%%%%%%%%%%%%%%%%%%%

\paragraph{On hardware.}

The classical algorithms are implemented using Python, and they run on a CPU using a 12th Gen Intel Core i7-1260P processor with 16.0 GB memory. We use the CPLEX Python package (and IBM CPLEX Optimization Studio for larger inputs).

The quantum algorithm modeled as the constrained quadratic model (CQM) is implemented using PyQUBO~\cite{zaman, tanahashi2019application}. It is solved with D-Wave's LeapHybridCQMSampler, a custom application for Leap's quantum-classical hybrid CQM solvers. The Leap hybrid solvers are proprietary classical-quantum hybrid solvers offered by D-Wave to solve large CQM-based formulated problems on quantum processing units (QPUs). Because the formulations are submitted to remote solvers, a wall clock measurement is not sufficient. In the experiments, the runtime of quantum algorithms is queried using Ocean API.

%%%%%%%%%%%%%%%%%%%%%%%%%%%%%%%%%%%%%%%%%%%%%%%%%%%%%%%%%%%%%

\subsection{Experimental results and analysis}

Each algorithm has been tested $5$ times for each input on all networks with the results displayed in the plots. We record the pre-processing time, the solver time, the number of variables, the number of constraints, and the quality of the solution for each test. The running time is measured in seconds. 

%%%%%%%%%%%%%%%%%%%%%%%%%%%%%%%%%%%%%%%%%%%%%%%%%%%%%%%%%%%%%

\subsubsection{Pre-processing time}

The preprocessing procedure converts the transportation networks to complete digraphs that our algorithmic formulations can work on. This procedure is performed on classical machines and the preprocessing time is shown in Table~\ref{tbl:pre}. It shows that even for a network with more than a thousand nodes and more than two thousand edges, the preprocessing time is within 10 seconds.

\begin{table}[h!]
\centering
\begin{tabular}{|c|c|}
\hline
Network & Time (s)\\ \hline \hline
Sioux Falls & 0.0412 \\ \hline
Eastern Massachusetts & 0.1414 \\ \hline
Anaheim & 1.3151 \\ \hline
Chicago & 8.4158 \\ \hline
Barcelona & 9.3551 \\ \hline
Winnipeg & 9.7158 \\ \hline
\end{tabular}
\caption{Pre-processing time for the networks}
\label{tbl:pre}
\end{table}

%%%%%%%%%%%%%%%%%%%%%%%%%%%%%%%%%%%%%%%%%%%%%%%%%%%%%%%%%%%%%

\subsubsection{The solvers' running time}

\begin{figure}[h!]
    \centering
    \includegraphics[width=0.7\linewidth]{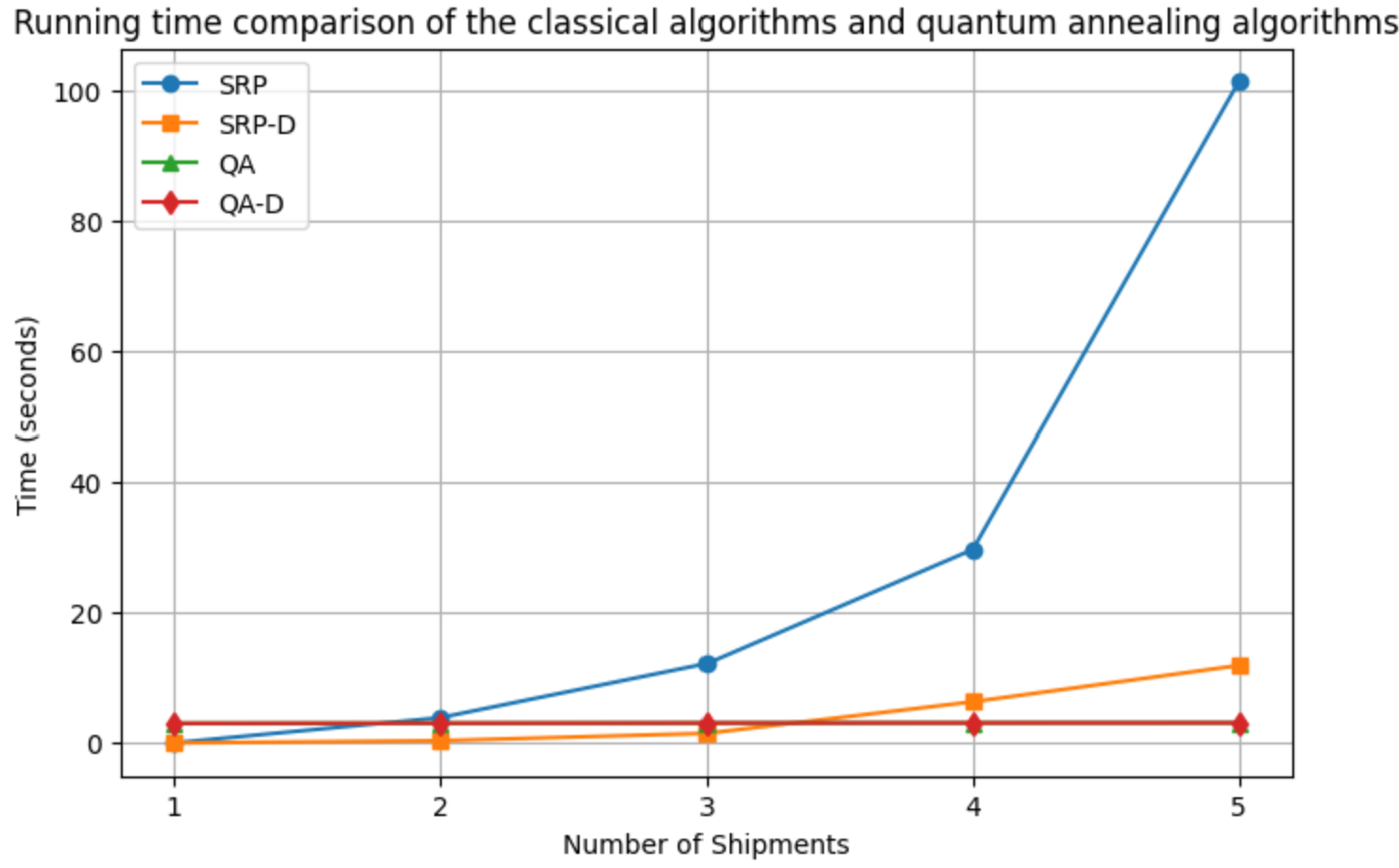}
    \caption{Running time comparison for the algorithms}
    \label{fig:compare}
\end{figure}

We show the running time of the classical algorithms in Table~\ref{tbl:classical}. We show the running time of the quantum annealing algorithms in Table~\ref{tbl:quantum} and we summarize the running-time comparison in Figure~\ref{fig:compare}.

\begin{table}[h!]
    \centering
    \begin{tabular}{|c|c|c|} \hline
    $\#$ of shipments & SRP's running time (s) & SRP-D's running time\\ \hline \hline
    1 & 0.0285 & 0.0160\\ \hline
    2 & 3.8760 & 0.3600\\ \hline
    3 & 12.1617 & 1.4839\\ \hline
    4 & 29.6739 & 6.3276\\ \hline
    5 & 101.2672 & 11.8923\\ \hline
    \end{tabular}
    \caption{CPLEX solving time for various $\#$ of shipments using SRP and SRP-D}
    \label{tbl:classical}
\end{table}

\begin{table}[h!]
    \centering
    \begin{tabular}{|c|c|c|}\hline
    $\#$ of shipments & QA's running time (s) & QA-D running time (s)\\ \hline \hline
    1 & 3.0264 & 3.0454\\ \hline
    2 & 3.0499 & 3.0786\\ \hline
    3 & 3.0794 & 3.1222\\ \hline
    4 & 3.0946 & 3.1682\\ \hline
    5 & 3.1024 & 3.2013\\ \hline
\end{tabular}
\caption{LeapHybridCQMSampler runtime for various $\#$ of shipments using QA and QA-D}
\label{tbl:quantum}
\end{table}

It has been observed that with the dispatch center in our formulation, the running times of the classical algorithm SRP-D are getting significantly smaller compared to those formulations without the dispatch center. The explanation is below: With the dispatch center conditions $c_\alpha$ and $c_\beta$, the CPLEX solvers limit the search space for the other conditions $C_1$ to $C_9$ and, thus, the running time gets smaller. This observation motivates us to think about how to integrate more conditions to enforce the existing algorithm solvers to search faster in the space.

For the shipment rerouting problem without and with a dispatch center, the running times of the quantum annealing algorithms QA and QA-D are very close to each other. An explanation for this observation is that the quantum annealing approaches generate the solutions by manipulating the qubits in parallel. The search space for the quantum annealing approaches does not depend on the sequential order of satisfying the conditions.

We observe that the quantum algorithms have higher running time overhead costs depending on the size of the graph. Figure~\ref{fig:compare} clearly indicates that the running times of the quantum annealing algorithms are increasing linearly, while the classical algorithms' running times are increasing exponentially. Part of the reason is that in the CQM formulation, the number of binary variables does not scale at the same rate as it does with the mixed-integer linear programs.
    
%%%%%%%%%%%%%%%%%%%%%%%%%%%%%%%%%%%%%%%%%%%%%%%%%%%%%%%%%%%%%

\subsubsection{The number of variables and constraints in the formulations}

\begin{figure}[h!]
    \centering
    \includegraphics[width=0.7\linewidth]{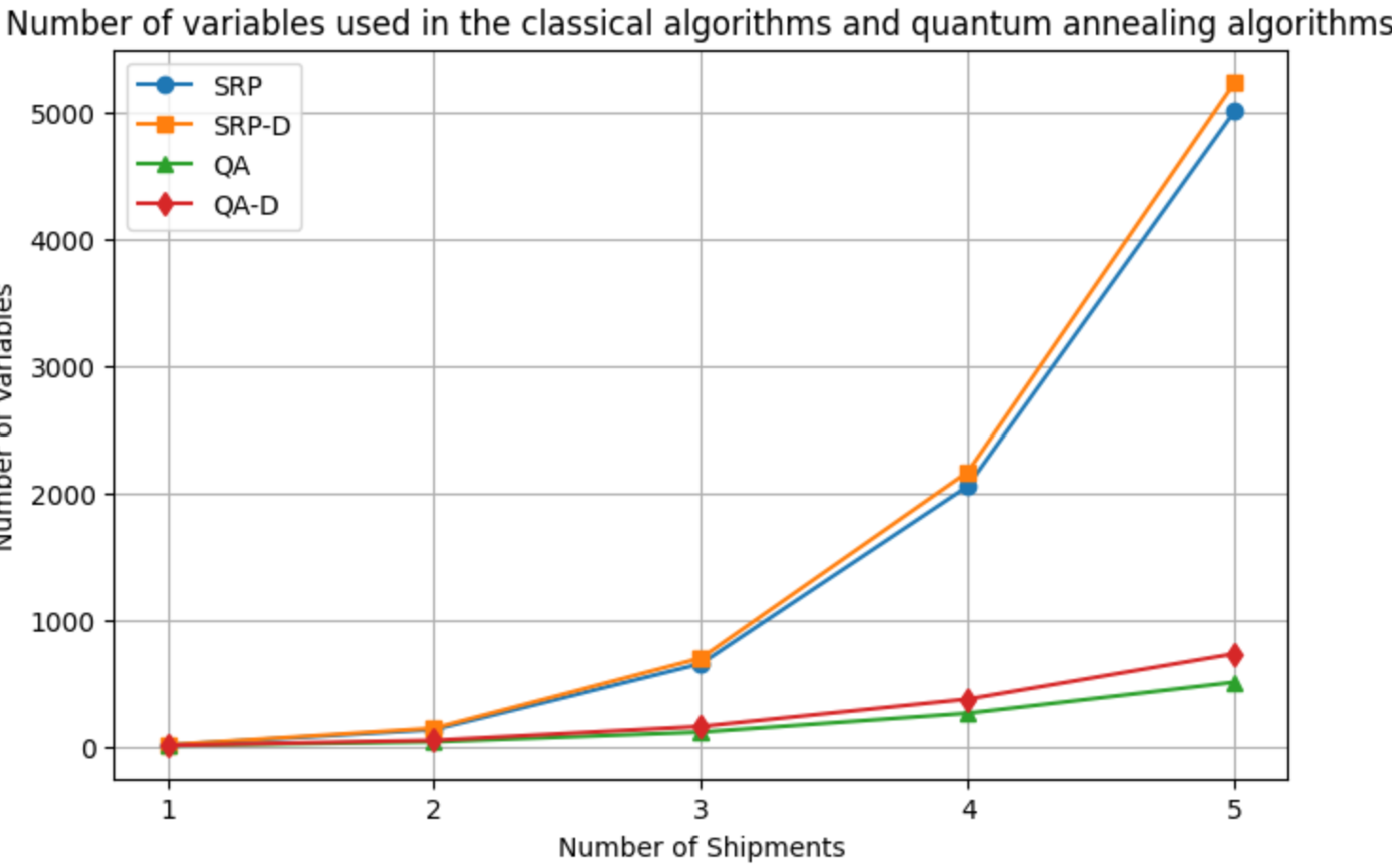}
    \caption{Relationship between number of shipments and the number of variables used in the algorithms}
    \label{fig:variables}
\end{figure}

\begin{figure}[h!]
    \centering
    \includegraphics[width=0.7\linewidth]{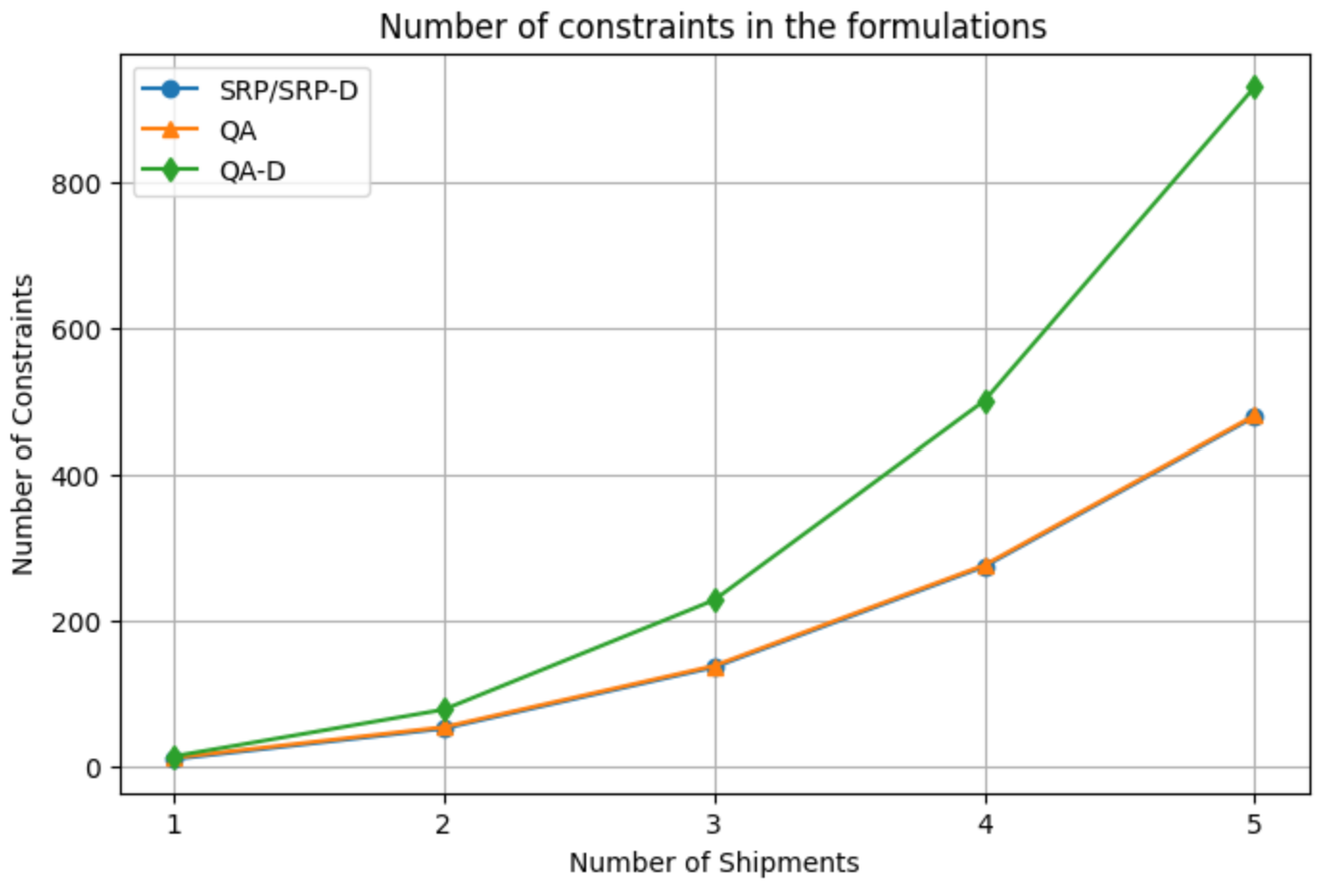}
    \caption{Relationship between number of shipments and the number of conditions required to be satisfied}
    \label{fig:constraints}
\end{figure}

Two key factors determining the running time of the algorithms are the number of variables and the number of constraints used in the formulations. We list the number of variables in Table~\ref{tbl:variables}, visualize them in Figure~\ref{fig:variables}, list the number of constraints in Table~\ref{tbl:constraints}, and visualize them in Figure~\ref{fig:constraints} respectively.

\begin{table}[h!]
    \centering
    \begin{tabular}{|c|c|c|c|c|}
    \hline
    $\#$ of shipments & SRP  & SRP-D & QA & QA-D\\ \hline \hline
    1     & 9    & 10     & 5  & 6\\ \hline
    2     & 130  & 142    & 34  & 46\\ \hline
    3     & 651  & 696    & 111  & 156 \\ \hline
    4     & 2052 & 2164   & 260  & 372 \\ \hline
    5     & 5005 & 5230   & 505  & 730\\ \hline
    \end{tabular}
    \caption{Relationship between number of shipments and the number of binary variables required}
    \label{tbl:variables}
\end{table}

\begin{table}[h!]
    \centering
    \begin{tabular}{|c|c|c|c|c|}
    \hline
    $\#$ of shipments & SRP & SRP-D  & QA & QA-D\\ \hline \hline
    1     & 10 & 10   & 12  & 14\\ \hline
    2     & 52 & 52   & 54  & 78\\ \hline
    3     & 136 & 136  & 138 & 228 \\ \hline
    4     & 274 & 274  & 276 & 500\\ \hline
    5     & 478 & 478  & 480 & 930\\ \hline
\end{tabular}
\caption{Relationship between number of shipments and the number of constraints.}
\label{tbl:constraints}
\end{table}

We observe that the quantum annealing algorithms' numbers of variables are much fewer than (about $1 / 6$ of) the classical algorithms. This observation also demonstrates the advantages of our quantum annealing approaches to solving the shipment rerouting problem. The numbers of conditions to be satisfied for both classical algorithms and quantum annealing algorithms are pretty close to each other.

%%%%%%%%%%%%%%%%%%%%%%%%%%%%%%%%%%%%%%%%%%%%%%%%%%%%%%%%%%%%%

\subsubsection{The algorithms' qualities}

\begin{figure}[h!]
    \centering
    \includegraphics[width=0.7\linewidth]{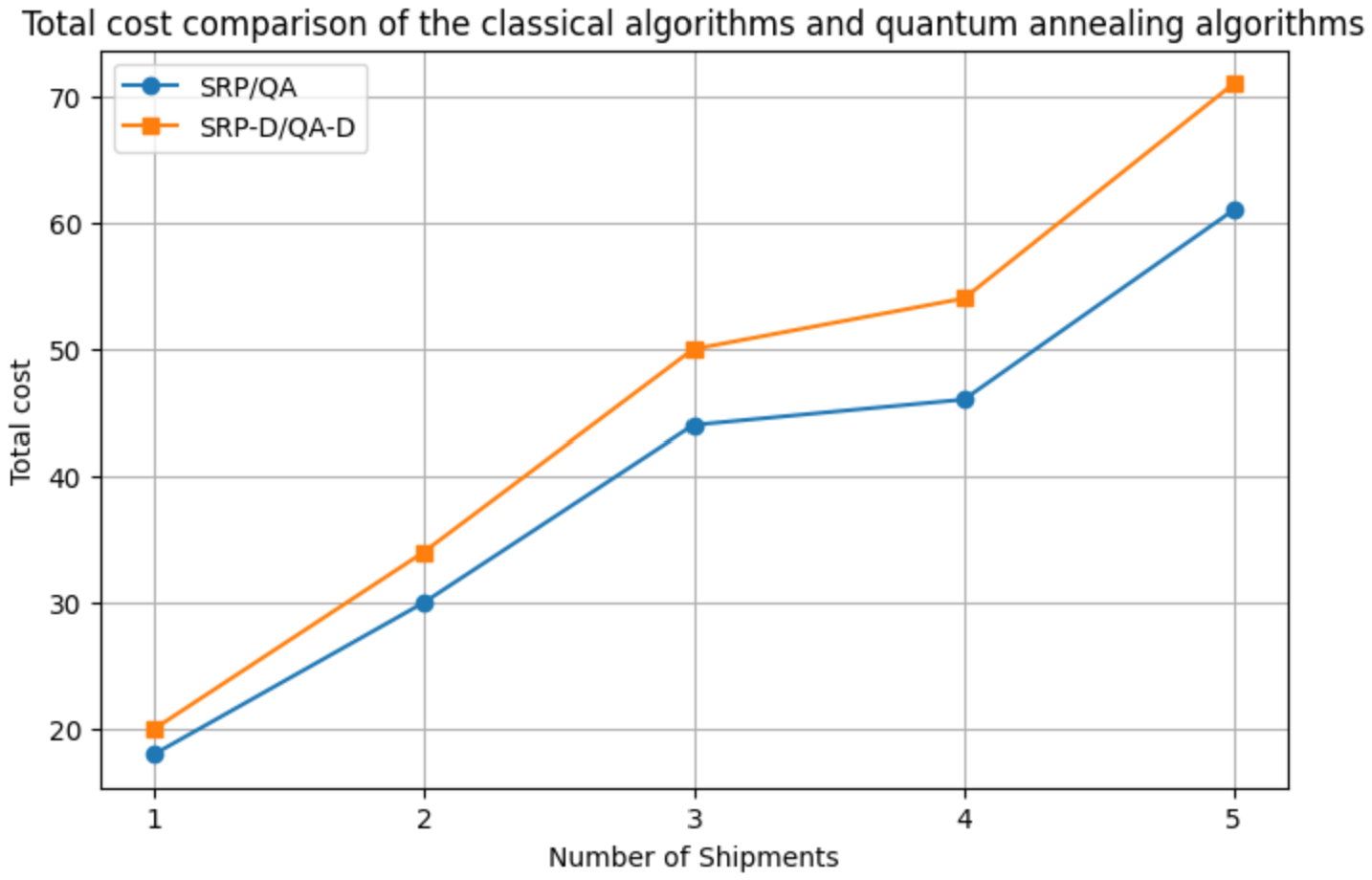}
    \caption{Relationship between number of shipments and the total cost}
    \label{fig:cost}
\end{figure}

We use the total cost to measure an algorithm's quality. In Table~\ref{tbl:cost}, we show the total cost of these algorithms on one network (Sioux Falls). We visualize the total cost in Figure~\ref{fig:cost}.

\begin{table}[h!]
    \centering
    \begin{tabular}{|c|c|c|c|c|}
    \hline
    $\#$ of shipments & SRP & SRP-D  & QA & QA-D\\ \hline \hline
    1     & 18.0 & 20.0 & 18.0 & 20.0\\ \hline
    2     & 30.0 & 34.0 & 30.0 & 34.0\\ \hline
    3     & 44.0 & 50.0 & 44.0 & 50.0\\ \hline
    4     & 46.0 & 54.0 & 46.0 & 54.0\\ \hline
    5     & 61.0 & 71.0 & 61.0 & 71.0\\ \hline
    \end{tabular}
    \caption{Relationship between number of shipments and the total cost (Sioux Falls)}
    \label{tbl:cost}
\end{table}

It is reasonable that with a dispatch center, the cost is getting more, for both the classical algorithms and the quantum annealing algorithms. Our observations indicate that the quantum annealing algorithms work optimally as the classical solutions. Due to their much less running time, we suggest applying the quantum annealing algorithms to solve the shipment rerouting problem.

%%%%%%%%%%%%%%%%%%%%%%%%%%%%%%%%%%%%%%%%%%%%%%%%%%%%%%%%%%%%%

\section{Conclusions}
\label{secc:conclusion}

%%%%%%%%%%%%%%%%%%%%%%%%%%%%%%%%%%%%%%%%%%%%%%%%%%%%%%%%%%%%%

In this paper, we consider the shipment rerouting problem. We give two classical algorithms and two quantum annealing algorithms. The next step of our research is to develop fast quantum algorithms for this problem. Motivated by the fact that the branch-and-bound quantum algorithms~\cite{Montanaro20, ChakrabartiMYP22} in some cases can yield running times that are substantially better than naively using Grover's algorithm, we design a variational quantum optimization algorithm for the branch-and-price framework. We are going to compare this quantum approximation optimization algorithm against the solutions proposed in this work. Also, based on our experiments, the running time is heavily dependent on the number of variables and the number of constraints in the formulations. We will work on formulations with fewer variables and constraints required. Motivated by the difference in the running time of the formulations with and without a dispatch center, we will also consider how to narrow down the search space by introducing more variables in the formulations.

%%%%%%%%%%%%%%%%%%%%%%%%%%%%%%%%%%%%%%%%%%%%%%%%%%%%%%%%%%%%%

\bibliographystyle{plain}
\bibliography{rerouting}

%%%%%%%%%%%%%%%%%%%%%%%%%%%%%%%%%%%%%%%%%%%%%%%%%%%%%%%%%%%%%

\end{document}